\documentclass[twocolumn,noshowpacs,prl,preprintnumbers,superscriptaddress,amsmath,amssymb,floatfix, groupedaddress]{revtex4}
\usepackage[caption=false]{subfig}
\usepackage{graphicx}
\usepackage{color}
\usepackage{setspace}
\usepackage{placeins}
\usepackage{tikz}
\usepackage{mathtools}
\usepackage{amsthm}
\usepackage{color}
\usepackage{xkeyval,xcolor}

\newcommand{\er}{Erd\H{o}s-R\'{e}nyi}

\DeclareFontFamily{OT1}{pzc}{}
\DeclareFontShape{OT1}{pzc}{m}{it}{<-> s * [1.10] pzcmi7t}{}
\DeclareMathAlphabet{\mathpzc}{OT1}{pzc}{m}{it}

\begin{document}

\title{Two transitions in spatial modular networks}
\author{Bnaya Gross} 
\affiliation{Bar-Ilan University, Ramat Gan, Israel}
\author{Dana Vaknin}
\affiliation{Bar-Ilan University, Ramat Gan, Israel}
\author{Sergey V.Buldyrev}
\affiliation{Yeshiva University, New York, USA}
\author{Shlomo Havlin}
\affiliation{Bar-Ilan University, Ramat Gan, Israel}
\affiliation{Institute of Innovative Research, Tokyo Institute of Technology, Midori-ku, Yokohama, Japan 226-8503}
\date{\today}

\begin{abstract}
	Understanding the resilience of infrastructures such as transportation network has significant importance for our daily life. Recently, a homogeneous spatial network model was developed for studying spatial embedded networks with characteristic link length such as power-grids and the brain. However, although many real-world networks are spatially embedded and their links have characteristics length such as pipelines, power lines or ground transportation lines they are not homogeneous but rather heterogeneous. For example, density of links within cities are significantly higher than between cities. Here we present and study numerically and analytically a similar realistic heterogeneous spatial modular model using percolation process to better understand the effect of heterogeneity on such networks. The model assumes that inside a city there are many lines connecting different locations, while long lines between the cities are sparse and usually directly connecting only a few nearest neighbours cities in a two dimensional plane. We find that this model experiences two distinct continuous transitions, one when the cities disconnect from each other and the second when each city breaks apart. Although the critical threshold for site percolation in 2D grid remains an open question we analytically find the critical threshold for site percolation in this model. In addition, while the homogeneous model experience a single transition having a unique phenomenon called \textit{critical stretching} where a geometric crossover from random to spatial structure in different scales found to stretch non-linearly with the characteristic length at criticality. Here we show that the heterogeneous model does not experience such a phenomenon indicating that critical stretching strongly depends on the network structure.
\end{abstract}
\maketitle

In the past two decades network science has provided a unique tool for analyzing complex systems by introducing novel frameworks for predicting and understanding collective phenomena in various systems such as the brain \cite{moretti2013griffiths_critical_brain,sporns2010networks_brain}, climate networks \cite{yamasaki2008climate,jingfan2017network_climate,ludescher2014very_climate}, epidemic spreading \cite{pastorsatorras-prl2001,gallos2012collective_obesity} and infrastructures \cite{yang2017small_power_grid,latora2005vulnerability_infrastructures,li-pnas2015}. One of the main tools used in networks science for describing functionality of a complex system is percolation theory \cite{staufferaharony,bunde1991fractals}. Percolation theory describes a physical process in which one randomly removes a fraction $1-p$ of nodes or edges from the network and analyses the network behaviour under such removal. At a critical point $p_c$ the system breaks apart and only small clusters remains. Thus, the system's functionality is usually described by the giant component which exist above criticality, $p_c$, and not below. \par

One of the most important properties of a network is its structure. The structure of a network can vary in many ways such as interplay between random to spatial structure \cite{danziger-epl2016,grossvaknin-JPS,barthelemy2003crossover_scalefree_spatial,vaknin-njp2017,bonamassagross2019critical}, different degree distribution \cite{barabasi-science1999}, clustering \cite{watts-nature1998,amaral-pnas2000} or community structure \cite{capocci2005detecting,shekhtman2015resilience_communities,newmanngirvan2002community,gross2019interconnections} which have a significant effect on the phenomena that appear on it. \par

Recently, a homogeneous spatial network model was developed for studying spatial embedded networks with characteristic link length such as power-grids, transportation systems and the brain. However, although many real-world networks have spatial embedding and a characteristics length such as pipelines, power lines or ground transportation lines \cite{danziger-epl2016} they are not homogeneous but rather heterogeneous since within  the same city lines are dense and connect arbitrary locations, while there are few lines between different cities, which are embedded in two dimensional space and usually limited to a few nearest neighbours. Therefore a heterogeneous spatial modular network model with characteristic link length is needed to better understand the effect of heterogeneity on such networks. \par

\begin{figure}
	\centering
	\begin{tikzpicture}[      
	every node/.style={anchor=north east,inner sep=0pt},
	x=1mm, y=1mm,
	]   
	\node (fig3) at (0,0)
	{\includegraphics[scale=0.45]{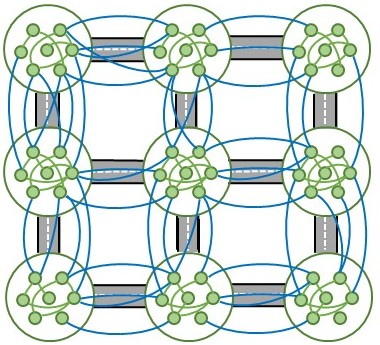}};
	\end{tikzpicture}
	\caption{\textbf{Illustration of the model.} The heterogeneous spatial modular model represents a structure of a network inside cities and between cities. Inside a city it is easy the get from one place to another (green links) like \er~network having random like structure while travelling from one city to another is usually possible between neighbouring cities having spatial like structure (blue links).}
	\label{fig:Illustration}	
\end{figure}

In this paper we study a  heterogeneous spatial modular model in 2D \cite{vaknin2019spreading}, see Fig. \ref{fig:Illustration}, to better describe heterogeneous infrastructures and study its resilience numerically and analytically using percolation theory. We find that such networks experience two phase transitions one at $p^{spatial}_c$ when the cities start to become disconnected from one another and the second at $p^{ER}_c$ when each city breaks apart itself. Although the critical threshold for site percolation in 2D grid is still an open question and is found only numerically \cite{newman2000efficient_site_perc}, here we find analytically $p^{spatial}_c$ for site percolation for this model by utilizing the well-known threshold of bond percolation in 2D. The two transitions are in contrast with a similar but homogeneous model \cite{danziger-epl2016,grossvaknin-JPS,vaknin-njp2017,bonamassagross2019critical} where a single  $p_c$ is found.

In addition, a new phenomena called \textit{critical stretching} has been found in the homogeneous model \cite{bonamassagross2019critical}. In this model a geometric crossover from mean-field behaviour in small scales to spatial in large scales has been observed. It was found that at criticality the mean-field regime stretches non-linearly with the characteristic scale. This model is similar to ours since both have mean-field behaviour on small scales and spatial on large scales. However, they are different in the assignment of the links. Our model has heterogeneous structure since it is divided into communities or cities whereas the model in  \cite{bonamassagross2019critical} has an homogeneous structure. Thus, we will refer to them as the heterogeneous model here and the homogeneous model in \cite{bonamassagross2019critical}. We find that while in the homogeneous structure one observes critical stretching, in contrast, the heterogeneous structure studied here, do not experience such a phenomena due to the two distinct transitions. \par


\underline{\textit{Heterogeneous and homogeneous spatial models.--}} The heterogeneous spatial modular model \cite{vaknin2019spreading} represents an infrastructure as a 2-dimensional square lattice with $N=L \times L$ lattice sites, where $L$ is linear size of the lattice. We assume that the lattice sites are the nodes of the heterogeneous network. However, the edges of the heterogeneous nodes do not coincide with the lattice bonds. The lattice is divided into a smaller squares of linear size $\zeta$ representing communities, e.g., cities. The number nodes in each community is $N_c= \zeta \times \zeta$. Thus, the number of communities in our model is $n = N/N_c=L^2 / \zeta^2$. We assume that inside a community it is easy to get from one node to another and therefore each community will be connected randomly as an \er~network (ER) with average degree $k_{intra}$. In contrast, it is difficult to get from one community to the next. Thus, we assume that in addition to intra-links linking the nodes in the same community there are much fewer inter-links which connect the nodes located in neighbouring communities. We assume that each node has inter-links distributed according to a Poisson distribution with the average degree $k_{inter}$.  Each community has four nearest neighbouring communities occupying adjacent squares on the lattice. We assume that the inter-links emanating from a community connect it randomly selected nodes in the four neighbouring communities. This assumption represents the fact that roads or railways connect neighbouring cities. For brevity of notations we denote $K\equiv k_{intra}$ and $Q\equiv k_{inter}\zeta^2$, where $Q$ is the average number of inter-links emanating from each community. \par
The homogeneous model studied in  \cite{danziger-epl2016,grossvaknin-JPS,vaknin-njp2017,bonamassagross2019critical} assumes 2-dimensional grid size $L \times L$ with $L$ being the lattice length. The construction of the model consists of the following stages: $i)$ A single node is randomly chosen. $ii)$ An edge length is randomly drawn from an exponential distribution $\mathcal{P}(r) \sim \exp(-\zeta / r)$ where $\zeta$ is the characteristic length of the distribution. $iii)$ The node is being connected to a random node at the distance of the drawn edge in $ii)$. These stages are repeated until the average degree $K$ is achieved. \par
These two models have the same two important limits. For $\zeta \to L$ the models generate an ER while for $L \gg \zeta \to 0$ strong spatial (regular lattice) behaviour is observed. Moreover, for intermediate values of $L > \zeta > 0$ mean-field behaviour is observed in small scales and spatial behaviour on large scales. The main difference between the models is the structure, heterogeneous versus homogeneous and as we show below in these intermediate scales (between mean-field and spatial) they behave very differently. 

\begin{figure}
	\centering
	\begin{tikzpicture}[      
	every node/.style={anchor=north east,inner sep=0pt},
	x=1mm, y=1mm,
	]   
	\node (fig3) at (0,0)
	{\includegraphics[scale=0.33]{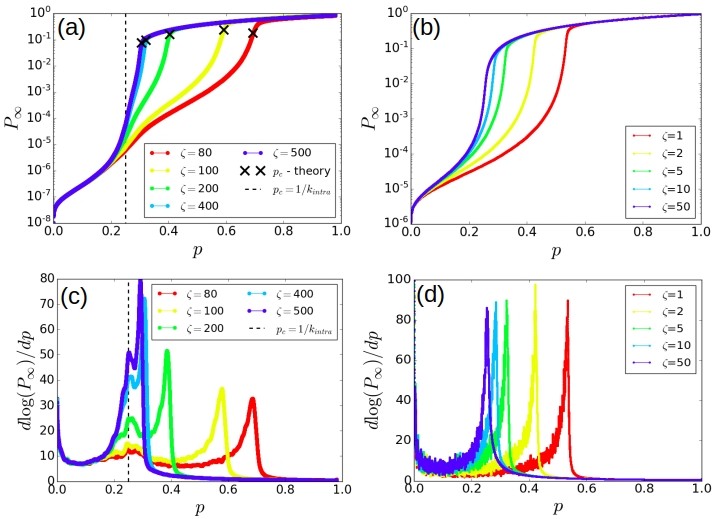}};
	\end{tikzpicture}
	\caption{\textbf{Simulations of the largest cluster $P_{\infty}$ as a function of $p$ for different values of $\zeta$ on semi log scale with $K = 4$ and $k_{inter} = 10^{-3}$. (a) The heterogeneous model.} Two distinct transitions are observed. The first (higher) transition at $p^{spatial}_c$ of the lattice is obtained from Eq. \eqref{eq_pc} and is denoted by black $\times$. The second (lower) transition occurs when the small ER communities break apart at $p^{ER}_c = 1/ K$. \textbf{(c)} Both transitions can be clearly seen in the maximal values of  the derivative of the giant component. \textbf{(b) The homogeneous model with $K = 4$.} Only one transition is observed in contrast to the heterogeneous model. \textbf{(d)} The single transition (single maximal values) can be clearly seen in the derivative.}
	\label{fig:giant_zeta_single_layer}	
\end{figure}

\underline{\textit{Analytical and numerical results of percolation in the }} \underline{\textit{heterogeneous  model.--}} The homogeneous model has been studied in \cite{bonamassagross2019critical} so here we will study the heterogeneous spatial modular model. We study the model under percolation process at which a fraction $1-p$ of nodes are randomly removed from the network. The behaviour of the largest cluster size, $P_{\infty}$, as function of $p$ for different values of $\zeta$ but fixed $k_{inter}=0.001$ and fixed lattice size $L=10^4$ is shown in Fig. \ref{fig:giant_zeta_single_layer}a. As expected, for $\zeta \to L$ the behaviour of the network approaches the behaviour of a regular ER with $p^{ER}_c = 1 / K = 1/4$. It can be seen that for any value of $\zeta \gg 1$ the graph $P_\infty(p)$ has two inflection points corresponding to the two maxima of the slope $d\ln{P}_{\infty}/dp$ in Fig. \ref{fig:giant_zeta_single_layer}c. While the position of the maximum near $p=p^{ER}_c$ does not depend on $\zeta$, the position of the second (larger) one decreases with $\zeta$, and at large $\zeta$ it almost coalesces with the first one. As we will see, this second maximum corresponds to the bond percolation threshold of the network of communities, which has a topology of the square lattice. Near this maximum the network breaks into individual communities, isolated from each other, so the giant component of the entire network disappears but each community remains well connected and their largest clusters remain of the order of $\zeta^2$. We will call these clusters local giant components. Finally, near the first (lower $p_c$) maximum, corresponding to the percolation threshold of the ER the local giant components disappear as well and the average largest cluster size swiftly goes to zero as $p$ decreases below $p^{ER}_c$. This behaviour is in marked contrast to the homogeneous model which experience a single transition threshold for any value of $\zeta$ (Fig. \ref{fig:giant_zeta_single_layer}b,d).

To demonstrate that the second inflection point (at higher $p$) corresponds to the bond percolation transition on the square lattice, we compute the position of the inflection points for different $\zeta$ analytically using the well know fact that the bond-percolation threshold for a square lattice, $p_b = 1/2$ \cite{bunde1991fractals}. Here we will use $p_b$ to find the value of $p^{spatial}_c$ at which the communities disconnect from each other.  The probability that one of $Q$ interlinks emanating from a given community connects to one of its 4 neighbours is $1/4$. Therefore, the number $k$ of the interlinks connecting these two neighbouring communities is distributed with a binomial distribution $P_k(Q)=(1/4)^k(3/4)^{Q-k}C_Q^k$. The probability that a randomly chosen node will be connected to the local giant component of each community is given by the giant component of ER network \cite{staufferaharony,erdHos1960evolution_random_grpahs},
\begin{equation}
G = p(1 - \exp^{-KG)}) .
\label{eq_ER}
\end{equation}
For small $k_{inter}$, the communities are weakly connected to each other similarly to weakly interacting networks \cite{rapisardi2019fragility}. In order for an inter-link connecting two neighbouring communities to  participate in building the  global giant component of the network, its both ends should belong to the local giant components of these two communities. This is the finite cluster of size $s\ll \zeta^2$ within a community will have a very low chance to have more than one interlink if $s \cdot k_{inter}\ll 1$. However, finite clusters within communities can still participate in the global connectivity once $k_{inter}$ is large enough and their size $s$ is such that $s \cdot k_{inter} \approx 1$. Thus, assuming small $k_{inter}$, the probability that an inter-link participates in the global connectivity is $G^2$ and the probability that two neighbouring communities will not have a single inter-link connecting  their local giant components is
\begin{equation}
p_b=\sum_k P_k(Q)(1-G^2)^k=[3/4+1/4(1-G^2)]^Q .
\label{eq_pb}
\end{equation}
At the lattice percolation threshold, the  probability that two neighboring communities do not have a bond connecting them, $p_b$ should be $1/2$, the bond percolation threshold. In this case, the size of each surviving community, $G(p^{spatial}_c)$, is large since $p^{ER}_c < p^{spatial}_c$ and each community itself is above criticality. $G(p^{spatial}_c)$ can be found analytically directly from Eq. \eqref{eq_pb},
\begin{equation}
G(p^{spatial}_c) = 2\sqrt{1-2^{-1/Q}} ,
\label{eq_G_pc}
\end{equation}
and the percolation threshold $p^{spatial}_c$ where the communities disconnect from one another can be obtained using Eq. \eqref{eq_ER} and \eqref{eq_G_pc},
\begin{equation}
p^{spatial}_c = \frac{2\sqrt{1-2^{-1/Q}}}{1 - \exp(-2K\sqrt{1-2^{-1/Q}})}.
\label{eq_pc}
\end{equation}
Indeed, in the limit of $\zeta \to \infty$, Eq. \eqref{eq_G_pc} takes the form
\begin{equation}
G(p^{spatial}_c)  \simeq \frac{2}{\zeta}\sqrt{\frac{\ln 2}{k_{inter}}} \sim \frac{1}{\zeta \sqrt{k_{inter}}},
\label{eq_G_pc_approx}
\end{equation}
and $p^{spatial}_c = p^{ER}_c = 1/K$. \par
\begin{figure}
	\centering
	\begin{tikzpicture}[      
	every node/.style={anchor=north east,inner sep=0pt},
	x=1mm, y=1mm,
	]   
	\node (fig3) at (0,0)
	{\includegraphics[scale=0.27]{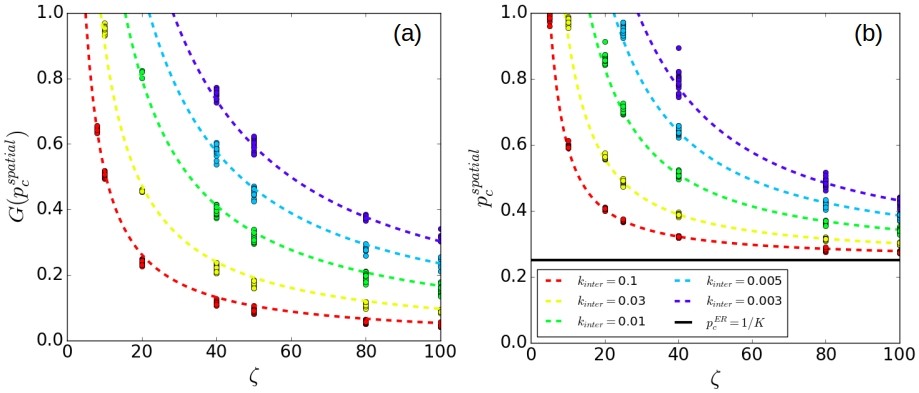}};
	\end{tikzpicture}
	\caption{\textbf{(a) $G(p^{spatial}_c)$ and (b) $p^{spatial}_c$ as a function of $\zeta$ for various values of $k_{inter}$ with $K = 4$.} The circles represent simulation results and the dashed lines are the theory obtained from \textbf{(a)} Eq. \eqref{eq_G_pc} and \textbf{(b)} Eq. \eqref{eq_pc} . In the limit of $\zeta \to \infty$ the system approaches a single ER network and $p^{spatial}_c$ approaches $p^{ER}_c = 1/K$.}
	\label{fig:pc_zeta_single_layer}	
\end{figure}

Simulation and theory for $p^{spatial}_c$ for different values of $\zeta$ and $k_{inter}$  are shown in Fig. \ref{fig:pc_zeta_single_layer}. The theory is obtained directly from Eq. \eqref{eq_pc} is seen to be in excellent agreement with simulations. In the limit $\zeta \to  \infty$ the network approaches the behaviour of a regular ER and Eq. \eqref{eq_pc} yields $p^{spatial}_c = p^{ER}_c = 1/K$. \par
\begin{figure}
	\centering
	\begin{tikzpicture}[      
	every node/.style={anchor=north east,inner sep=0pt},
	x=1mm, y=1mm,
	]   
	\node (fig3) at (0,0)
	{\includegraphics[scale=0.31]{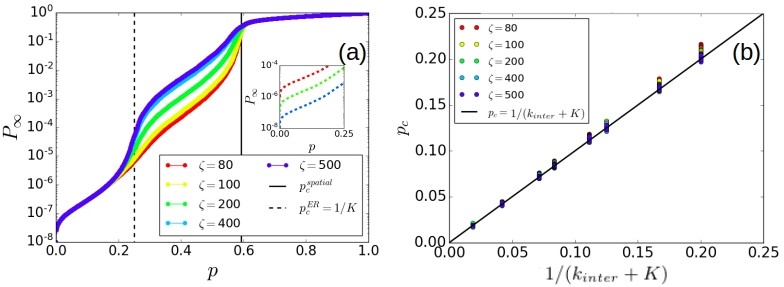}};
	\end{tikzpicture}
	\caption{\textbf{(a) The giant component $P_{\infty}$ as a function of $p$ for different values of $\zeta$ on semi log scale with $K = 4$ and $Q = 10$.} $p^{spatial}_c$ depends only on $Q$ (Eq. \eqref{eq_pc}). The inset shows finite size effect in the $p < p^{ER}_c (=0.25)$ regime. Indeed, as $N$ increases $P_{\infty}$ goes to zero ($N = 10^6, 10^7, 10^8$ - red, green and blue respectively) in this regime. \textbf{(b) $p^{spatial}_c$ as a function of $1/(k_{inter} + K)$ for $K = 4$ and large values of $k_{inter}$ for different values of $\zeta$.} For large values of $\zeta$ the network is similar to ER networks and therefor $p^{spatial}_c = 1/(k_{inter} + K)$. Here $L = 10^4$.}
	\label{fig:pc_zeta_single_layer_large_kinter}	
\end{figure}

The percolation threshold $p^{spatial}_c$ depends only on the average inter-degree of a city $Q$ as predicated by Eq. \eqref{eq_pc}.
In Fig. \ref{fig:pc_zeta_single_layer_large_kinter}a we show $P_{\infty}$ for different values of $\zeta$ while keeping $Q$ fixed. Indeed, as predicated, $p^{spatial}_c$ does not change. However, as $\zeta$ increases while $Q$ is fixed in the interval of $p^{ER}_c<p<p_c^{spatial}$ the sizes of local giant components are increasing as $\zeta^2$. Thus, although the global giant component does not exist in this interval, the fraction of nodes in the largest cluster grows as $\zeta^2/L^2$. In addition, when $\zeta$ becomes comparable to $L$ the lattice of communities becomes very small and the spatial transition becomes less pronounced due to finite size effects. In the $p < p^{ER}_c$ regime a non-vanishing largest cluster seem to appear. However, this only due to finite size effect. As $N$ increases $P_{\infty} \to 0$ as seen in the inset of Fig. \ref{fig:pc_zeta_single_layer_large_kinter}a.\par

So far, the theoretical analysis assumed realistic small $k_{inter}$ which means that the communities are weakly connected to each other. This leads to the spatial transition described by Eqs. \eqref{eq_ER} and \eqref{eq_pb} when the communities become disconnected from each other and then their local giant components disappear as $p$ further decreases at $p^{ER}_c = 1/K$. However, for large $k_{inter}$ this description is not valid since the communities are strongly connected to each other and therefore even small clusters within the communities will participate in the global connectivity. Thus, the transition point where the communities disconnect from each other is the same as when each community breaks. The formalism presented in Ref. \cite{vaknin2019spreading} shows that for ER local networks $k_{inter}$ simply adds the $K$ in Eq. (\ref{eq_ER}) giving $p^{ER}_c = 1/(K + k_{inter})$ as shown in Fig. \ref{fig:pc_zeta_single_layer_large_kinter}b.

\begin{figure}
	\centering
	\begin{tikzpicture}[      
	every node/.style={anchor=north east,inner sep=0pt},
	x=1mm, y=1mm,
	]   
	\node (fig3) at (-45,0)
	{\includegraphics[scale=0.35]{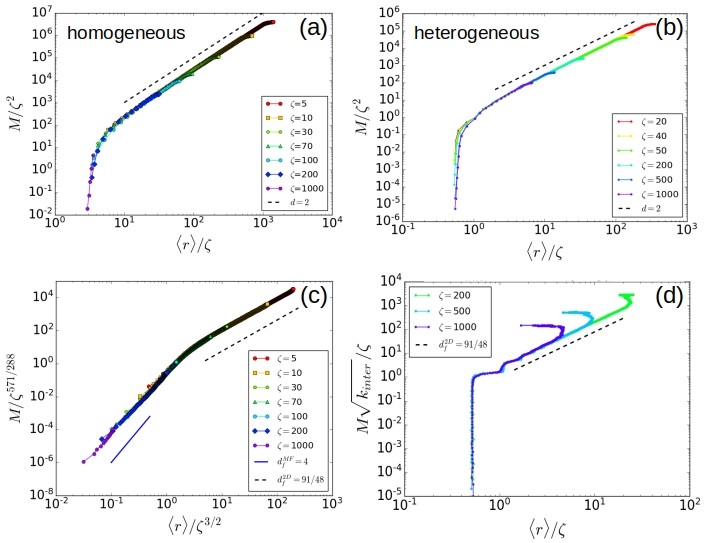}};
	
	\end{tikzpicture}
	\caption{\textbf{Comparing the geometric crossover between the homogeneous and heterogeneous models in $2D$ with $K = 4$.} The mass scaling at $p = 1$ for the \textbf{(a)} homogeneous model and \textbf{(b)} the heterogeneous model with $k_{inter} = 10^{-1}$. Both large and small scales are not at criticality leading to a similar geometric crossover at $r_*(\zeta) \equiv \zeta$ from infinite dimension in small scales (small world) to $2D$ in large scale. \textbf{(c)} The critical stretching phenomena in the homogeneous model at $p = p_c$. Both large and small scales are at criticality and the random structure in small scales stretches non-linearly with $\zeta$ as $r_*(\zeta) \equiv \zeta^{3/2}$. A geometric crossover is observed at $r_*(\zeta)$ from $d^{MF}_f = 4$ in small scale to $d^{2D}_f = 91/48$ in large scale since both large and small scales are at criticality. \textbf{(d)} Absence of critical stretching phenomena in the heterogeneous model at $p = p^{spatial}_c$ with $k_{inter} = 10^{-4}$. Since $p^{spatial}_c > p^{ER}_c$ only large scales is at criticality leading to a geometric crossover at $r_*(\zeta) \equiv \zeta$ from infinite dimension in small scales to $d^{2D}_f = 91/48$ in large scales without critical stretching. Here $L = 10^4$.}
	\label{fig:dimension}	
\end{figure}

\underline{\textit{Geometric crossover and stretching.--}} The critical stretching phenomena introduced in Ref. \cite{bonamassagross2019critical} is observed at the critical percolation threshold of the homogeneous model. Since the homogeneous model has only one threshold (Fig. \ref{fig:giant_zeta_single_layer}b,d), both the random structure in small scale and the spatial structure in large scale are at criticality. This leads to two different fractal dimensions in both small and large scales with a geometric crossover at $r_*(\zeta)$ from $d^{MF}_f = 4$ for $r < r_*(\zeta)$ to $d^{2D}_f = 91/48$ for $r > r_*(\zeta)$ \cite{bunde1991fractals} where $r_*(\zeta)$ is the crossover point. The Ginsburg criterion \cite{ginzburg1961} can be applied to identify the crossover point and is given by $\mathpzc{p}^{3-d/2}\zeta^d \gg 1$ where $\mathpzc{p} = p/p_c - 1$ is the displacement of $p$ from its critical value. In $2D$ the crossover should occur at $\mathpzc{p}_* \equiv \zeta^{-1}$ which leads to $r_*(\zeta) \equiv \zeta^{3/2}$ as described in detail in Ref. \cite{bonamassagross2019critical}. This non-linear dependence of $r_*(\zeta)$ with $\zeta$ is the critical stretching phenomena which means that the random structure at small scales stretches at criticality in a non-linear fashion with $\zeta$. The mass of the giant component  $M_*(\zeta)$ at $\mathpzc{p}_*$ can also evaluated using consistency arguments giving $M_*(\zeta) \equiv \zeta^{571/288}$. Thus, we can write the ansatz,
\begin{equation}
\frac{M(r,\zeta)}{M_*(\zeta)} = \bigg(\frac{r}{r_*(\zeta)} \bigg)^{4}\mathcal{M}\bigg(\frac{r}{r_*(\zeta)} \bigg)
\label{stretching_homogeneous}
\end{equation}
where $\mathcal{M}(x) \propto x^{-101/48}$ for $x \gg 1$ and constant otherwise.
The scaling relation in Eq. \eqref{stretching_homogeneous} is supported by simulation presented in Fig. \ref{fig:dimension}c showing that indeed the random structure at small scales stretches at criticality. \par
In contrast to the above discussion, at $p = 1$, the homogeneous model neither in small nor in large scales is at criticality. Thus, a geometric crossover is expected and observed from infinite dimension in small scales (small world) to 2-dimension in large scales. The crossover point $r_*(\zeta) \equiv \zeta$ scales linearly with $\zeta$ and no stretching is observed as seen in Fig. \ref{fig:dimension}a. The mass at the crossover point scales as $M_*(\zeta) \equiv \zeta^2$. \par

In contrast to the homogeneous model, the heterogeneous model has two percolation transitions (Fig. \ref{fig:giant_zeta_single_layer}a,c). At $p^{spacial}_c$ only the large scale network is at criticality and the small scale networks are not since $p^{ER} < p^{spacial}_c$. Thus, we observe $2D$ fractal dimension at large scales and infinite dimension at small scales. Since small scales are not at criticality they do not stretch and $r_*(\zeta) \equiv \zeta$. The mass of the giant component at the crossover $M_*(\zeta)$ is the size of the local giant component at $p^{spacial}_c$. From Eq. \eqref{eq_G_pc_approx} we obtain $M_*(\zeta) \sim G(p^{spacial}_c)\zeta^2  \sim \zeta / \sqrt{(k_{inter})}$ as supported by simulation in Fig. \ref{fig:dimension}d. \par

At $p = 1$ both large and small scales are not at criticality similar to the homogeneous model. We observe a geometric crossover at $r_*(\zeta) \equiv \zeta$ from infinite dimension at small scales to $2D$ at large scales without critical stretching as shown in Fig. \ref{fig:dimension}b. 

\underline{\textit{Summary and discussion--.}} In this work we introduced a spatial modular model for spatial heterogeneous infrastructure like a system of pipelines, power lines or ground transportation lines. In small scales this model has random ER-like behaviour since inside a city the lines may connect  arbitrary locations with fast connections, while on large scales the model has spatial behaviour because  the lines between cities usually connect only nearest neighbours embedded in two dimensional space. This model experience two distinct transitions, one at $p^{spatial}_c$ when the cities disconnects from one another and the other $p^{ER}_c$ when each city breaks apart.\par
We compared our model to a similar homogeneous model under percolation process. Away from criticality both models show a geometric crossover at $r^*(\zeta) = \zeta$ from infinite dimension to 2-dimension. However, since the homogeneous model has a single $p_c$, at $p = p_c$ both large and small scales are at criticality leading to a geometric crossover from  mean-field fractal dimension to 2D fractal dimension at $r^*(\zeta) = \zeta^{3/2}$ and critical stretching is observed. However, since in the heterogeneous model $p^{spatial}_c > p^{ER}_c$, at $p = p^{spatial}_c$ only large scales are at criticality and small scales do not. This leads to a geometric crossover from infinite dimension to 2D fractal dimension without critical stretching indicating that in order to observe critical stretching the system has to be at criticality in both large and small scales.

\underline{\textit{Acknowledgments.--}}  We thank the Italian Ministry of Foreign Affairs and International Cooperation jointly with the Israeli Ministry of Science, Technology, and Space (MOST); the Israel Science Foundation, ONR, the Japan Science Foundation with MOST, BSF-NSF, ARO, the BIU Center for Research in Applied Cryptography and Cyber Security, and DTRA (Grants no. HDTRA-1-14-1-0017 and HDTRA-1-19-1-0016) for financial support. D. V. thanks the PBC of the Council for Higher Education of Israel for the Fellowship Grant. S.V.B. acknowledges the partial support of this research through the B. W. Gamson Computational Science Center at Yeshiva University. 

\bibliographystyle{unsrt}
\bibliography{mybib}

\end{document}